\documentclass[12pt]{article}
\usepackage{amssymb}
\usepackage{scicite}
\usepackage{graphicx}
\usepackage{times}

\newenvironment{sciabstract}{\begin{quote} \bf}
{\end{quote}}

\newcounter{lastnote}
\newenvironment{scilastnote}{\setcounter{lastnote}{\value{enumiv}}\addtocounter{lastnote}{+1}\begin{list}{\arabic{lastnote}.}
{\setlength{\leftmargin}{.22in}}
{\setlength{\labelsep}{.5em}}}
{\end{list}}

\begin{document}

\title{Fluxonium: single Cooper pair circuit free of charge offsets}
\author{Vladimir E. Manucharyan,$^{1}$ Jens Koch,$^{1}$ Leonid I.\ Glazman,$%
^{1}$ Michel H.\ Devoret$^{1}$ \\
%EndAName
\\
{\normalsize {$^{1}$Departments of Physics and Applied Physics, Yale
University, New Haven, Connecticut 06520, USA}}\\
}
\date{}
\maketitle


%\baselineskip24pt


\begin{sciabstract}
The promise of single Cooper pair quantum circuits based on tunnel junctions
for metrology and quantum information applications is severely limited by
the influence of \textquotedblleft offset\textquotedblright\ charges --
random, slowly drifting microscopic charges inherent to many solid-state
systems. By shunting a small junction with the Josephson kinetic inductance
of a series array of large capacitance tunnel junctions, thereby ensuring
that all superconducting islands are connected to the circuit by at least
one large junction, we have realized a new superconducting artificial atom
which is totally insensitive to offset charges. Yet, its energy levels
manifest the anharmonic structure associated with single Cooper pair
effects, a useful component for solid state quantum computation.
\end{sciabstract}

Electric charge can be manipulated at the level of a single charge quantum~%
\cite{singlechargebook} in two types of superconducting circuits with
different topologies. The minimal example of the first type of circuit is
the Cooper pair box, which consists of an isolated superconducting electrode
(\textquotedblleft island\textquotedblright ) connected to a superconducting
reservoir on one side by a small tunnel junction, and on the other side by a
gate capacitance in series with a voltage source. The dynamics of the island
is described by two variables: the integer number of Cooper pairs occupying
the island and its conjugate, the $2\pi $-cyclic superconducting phase
difference between the island and the reservoir. The junction area must be
chosen sufficiently small such that the electrostatic energy of the island
due to an extra Cooper pair is larger than the Josephson energy of its
coupling to the reservoir, thus confining fluctuations of the number of
Cooper pairs below unity. Stated in electrical engineering language, one
needs $Z_{J}\gtrsim R_{Q}$, where the junction reactive impedance $%
Z_{J}=(L_{J}/C_{J})^{1/2}$ is defined by the Josephson characteristic
inductance $L_{J}$ and capacitance $C_{J}$ \cite{Josephson}, and where the
superconducting impedance quantum is given by $R_{Q}=\hbar /(2e)^{2}\approx
1~\mathrm{k\Omega }$, denoting Planck's constant $\hbar $ and the charge
quantum $e$. The second type of circuit is based on a superconducting loop
connecting the two electrodes of a small junction with an inductance which
exceeds $L_{J}$. The circuit conjugate variables are now the magnetic flux
generated by the persistent current in the loop and the displacement charge
on the plates of the small junction capacitance. When $Z_{J}\gtrsim R_{Q}$,
the large loop inductance is submitted to quantum fluctuations of flux
larger than the flux quantum $\Phi _{0}=2\pi \hbar /2e$, and therefore
according to Heisenberg principle, the junction charge fluctuations are
reduced below the value $2e$.

In practice, the realization of both circuit types faces fundamental
difficulties. Islands are exposed to random electric fields due to
fluctuating charged impurities which are ubiquitous in most solid-state
environments and whose compounded effect is described by a noisy offset
charge. Although the fully developed charging effects were demonstrated for
the Cooper pair box~\cite{BouchiatCPB,NakamuraCPB}, it soon became clear
that the low-frequency offset charge noise was a major source of decoherence
for charge qubits derived from this device~\cite%
{NakamuraCPB,SaclayQuantronium,ChalmersCPB,MetcalfeQuantroniumPRB}. This
state of affairs has prompted the development of alternative superconducting
qubits based on large junctions with $Z_{J}\ll R_{Q}$, avoiding the single
Cooper pair regime and the related charge offset problem~\cite%
{ChargeFreePhaseQubit,ChargeFreeFluxQubit,ChargeFreeTransmon}. On the other
hand, implementing the island-free circuit, which is immune to charge offset
noise, is another hard problem. This is because any finite-length wire with
inductance $L$ always comes with self-capacitance $C$ which reduces the
total charging energy of the circuit and therefore steers it away from the
charging regime, unless $(L/C)^{1/2}\gg R_{Q}$. In fact, a purely
electromagnetic inductance is incompatible with the single Cooper pair
effects since $(L/C)^{1/2}$ is then bounded by the vacuum impedance $(\mu
_{0}/\varepsilon _{0})^{1/2}\approx 377$~$\Omega <R_{Q}$, $\mu _{0}$ and $%
\varepsilon _{0}$ being vacuum permeability and permittivity~\cite%
{FeynmanVol2,FineStructureNote}.

In this paper, we present experimental results on a novel single Cooper pair
circuit based on a superconducting loop, which solves both the inductance
and the offset charge noise problems.

\begin{figure}[tbp]
\centering
\includegraphics[width = 1.0\columnwidth]{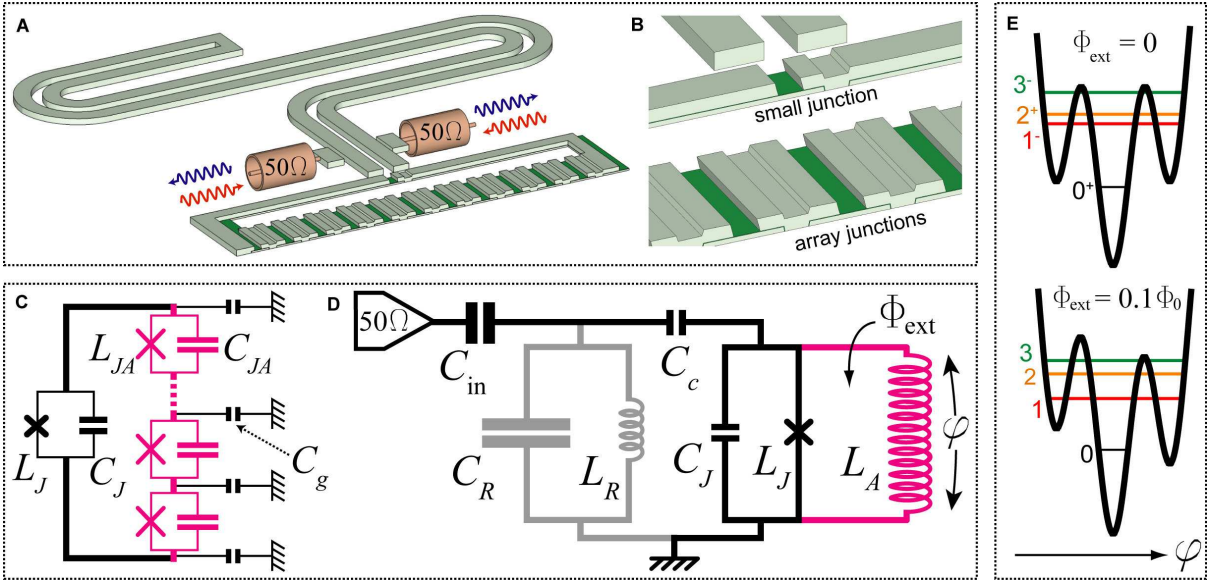}
\caption{(A) Sketch of a small Josephson junction shunted by an
array of larger area junctions. The two superconducting leads of the small
junction are coupled capacitively to a quarter-wave microwave resonator, a
\textquotedblleft parallel wire\textquotedblright\ transmission line shorted
on the opposite end. The resonator itself is probed capacitively and
symmetrically via two $50~\mathrm{\Omega }$ microwave ports, resulting in a
loaded quality factor of $400$. The whole device is made with a single step
standard Al/AlOx/Al double angle evaporation through an e-beam lithography
mask on a high resistivity Si substrate. (B) Close-up view of the small
junction region, showing top and bottom junction electrodes (grey) and their
thin oxide layer (green). Array junctions are about one order of magnitude
larger in area and spaced as tight as e-beam lithography resolution allows,
minimizing microwave parasitics. (C) Electrical circuit representation of
the loop formed by the small junction\ (black), with Josephson inductance $%
L_{J}$ and capacitance $C_{J}$, shunted by the array of larger junctions
(purple), with the corresponding inductance $L_{JA}$ and capacitance $C_{JA}$%
. Islands formed between the array junctions have small capacitance to
ground $C_{g}$. (D) Simplified circuit model of the fluxonium, consisting of
three sections: i) circuit equivalent to a Cooper pair box, where the small
junction with capacitance $C_{J}$ and non-linear Josephson inductance $L_{J}$
is capacitively, with capacitance $C_{c}$, coupled to the probe (solid black), such that $%
L_{J}/C_{J})^{1/2}>\hbar /(2e)^{2}$~; ii) giant inductance $L_{A}\gg L_{J}$
provided by the junction array (purple); iii) a parallel combination of $%
C_{R}$ and $L_{R}$ such that $(L_{R}/C_{R})^{1/2}\approx 50\Omega \ll \hbar
/(2e)^{2}$ which is the circuit model for distributed transmission line
resonator (grey). (E) Potential seen by the reduced flux $\varphi $ and
energy spectrum of the circuit (D) for two values of external flux $\Phi _{%
\mathrm{ext}}$. At $\Phi _{\mathrm{ext}}=0$, energy levels possess
well-defined parity as indicated with `+' and `-' signs next to level
numbers. Note that, in contrast with the RF-SQUID or flux qubit, there is on
average only one level per local minima.}
\label{fig:Fig1}
\end{figure}

The small junction of our circuit is shunted by a series array of carefully
chosen larger area tunnel junctions (Fig 1A-C). Here, all islands are
connected to the rest of the circuit by at least one large junction so that
quasistatic offset charges on all islands are screened. The large
capacitances of the array junctions prevent phase slips within the array,
and for excitations whose frequencies are below the junction plasma
frequency, the array effectively behaves as an inductive wire. By choosing a
sufficiently large number of array junctions it is possible to create an
inductance exceeding that of the small junction. At low energies, the loop
is effectively described by the loop flux $\hat{\Phi}$ and the small
junction charge $\hat{Q}$, satisfying $[\hat{\Phi},\hat{Q}]=i\hbar $.

To form a charge offset-free inductively shunted junction, four conditions
involving the effective inductance $L_{JA}$ and capacitance $C_{JA}$ of the $%
\mathcal{N}$ array junctions are required: (i) $\mathcal{N}L_{JA}\gg L_{J}$,
(ii) $e^{-8R_{Q}/Z_{JA}}<\varepsilon \ll 1$, (iii) $\mathcal{N}%
e^{-8R_{Q}/Z_{JA}}\ll e^{-8R_{Q}/Z_{J}}$, and (iv) $\mathcal{N<(}%
C_{JA}/C_{g})^{1/2}$. In the first relation (i), we simply estimate the
total array inductance to be $\mathcal{N}L_{JA}$ and require that it exceeds
the small junction inductance, allowing it to support the large flux
fluctuations of the loop. The second relation (ii), where $%
Z_{JA}=(L_{JA}/C_{JA})^{1/2}$ is the array junction reactive impedance,
dictates the minimum size of the array junctions necessary to reduce~\cite%
{JensTransmon} the uncontrolled offset charge on the islands of the circuit
below the desired value of the order of $2e\times \varepsilon $. The third
relation (iii) ensures that the inductive role of the array is not
jeopardized by quantum phase slips~\cite{GlazmanLarkinMatveev}.
Specifically, the probability amplitude of a phase slip event within the
array (l.h.s.) must be negligible compared to that in the small junction
(r.h.s.). According to relation (iii) a fluxon tunnels in and out of the
loop predominantly via the small junction, thus effectively erasing the
discrete character of the array. Lastly, relation (iv) states that the
inductance of the array is not shunted by the parasitic capacitances $C_{g}$
of array islands to ground. It is obtained by estimating the array parasitic
resonance frequency to be $(L_{JA}\mathcal{N}\times C_{g}\mathcal{N})^{-1/2}$%
, and requiring that it is larger than the junction plasma frequency $%
(L_{JA}C_{JA})^{-1/2}$. Remarkably, it is the relation (iv) which, with
present junction technology, most severely limits the maximum number of
junction in the array and, thus, its maximum inductance.

We have implemented the above array proposal and constructed a new
superconducting artificial atom which we have nicknamed \textquotedblleft
fluxonium\textquotedblright . It contains $\mathcal{N}=43$ Al-AlOx-Al
Josephson junctions~\cite{MaterialsAndMethods} such that $Z_{JA}\simeq
0.5R_{Q}$ and a small junction with $Z_{J}\simeq 1.5R_{Q}$~\cite%
{theEcfootnote}. The above four conditions being realized, the fluxonium can
be modelled (Fig.\ 1D) as a small junction shunted by an inductance $L_{A}$ 
\cite{footnote2}. The three characteristic energies of this model, namely $%
E_{L}=(\Phi _{0}/2\pi )^{2}/L_{A}$, $E_{J}=(\Phi _{0}/2\pi )^{2}/L_{J}$ and $%
E_{C}=e^{2}/\left( 2C_{J}\right) $ have values corresponding to $0.52~%
\mathrm{GHz}$, $9.0~\mathrm{GHz}$ and $2.5~\mathrm{GHz}$, respectively. The
additional $L_{R}C_{R}$ resonator, capacitively connected to the small
junction (Fig.\ 1D), reads out the \textquotedblleft atom\textquotedblright\
in a manner analogous to the dispersive measurement of cQED qubits~\cite%
{WallraffQED}. It is implemented by a quarter-wave superconducting coupled
microstrip resonator (Fig.\ 1A) with quality factor $400$ due to capacitive
coupling to the two $50~\mathrm{\Omega }$ measurement ports. The resonator
frequency $\omega _{R}=(L_{R}C_{R})^{-1/2}\simeq 2\pi \times 8.17~\mathrm{GHz%
}$ is pulled by the reactance of the fluxonium circuit and is monitored by
standard ultra low noise microwave reflection technique. The fluxonium
reactance depends on its quantum state, an effect leading to a purely
dispersive state measurement~\cite{MaterialsAndMethods}. An externally
imposed, static magnetic flux $\Phi _{\mathrm{ext}}$ threading the loop $%
\Phi _{0}$-periodically modulates the spacings of energy levels of our
artificial atom.

Introducing the operators $\hat{N}=\hat{Q}/2e$ and $\hat{\varphi}=2e\hat{\Phi%
}/\hbar $, describing the reduced charge on the junction capacitance and its
conjugate reduced flux operator~\cite{DevoretQFinEC}, the Hamiltonian of the
fluxonium coupled to its readout resonator can be written as 
\begin{equation}
\hat{H}=4E_{C}\hat{N}^{2}+\frac{1}{2}E_{L}\hat{\varphi}^{2}-E_{J}\cos \left( 
\hat{\varphi}-2\pi \Phi _{\mathrm{ext}}/\Phi _{0}\right) +g\hat{N}\left( 
\hat{a}+\hat{a}^{\dag }\right) +\hbar \omega _{R}\hat{a}^{\dag }\hat{a}
\label{Ham}
\end{equation}%
Here $\hat{a}$ is the photon annihilation operator for the resonator, $g$ is
the atom-resonator coupling constant. The second term and the range of
definition of $\hat{\varphi}$ and $\hat{N}$, whose eigenvalues are here both
on the entire real axis, distinguishes the form of Hamiltonian (1) from that
of the Cooper pair box in cQED\ experiments \cite{WallraffQED}. There are
three important points to note concerning this Hamiltonian~\cite{JensCPBL}:
i) it is invariant under the transformation $\hat{N}\rightarrow \hat{N}+N_{%
\mathrm{offset}}\ $($N_{\mathrm{offset}}$ stands for offset charge value)
hence the \textquotedblleft charge-free\textquotedblright\ character of our
device; ii) it differs from that of the transmon~\cite{JensTransmon} since
offset charge influence is screened for all states, not just for the
low-lying states; iii) its second term, despite the fact that $E_{L}$ is the
smallest of the fluxonium energies, has a non-perturbative influence on the
full energy spectrum of this artificial atom, which presents strongly
anharmonic transitions \cite{footnote3} (Fig.\ 1E). Our experiment probes
these transitions by microwave spectroscopy, from which we infer the size of
charge fluctuations.

\begin{figure}[tbp]
\centering
\includegraphics[width = 0.5\columnwidth]{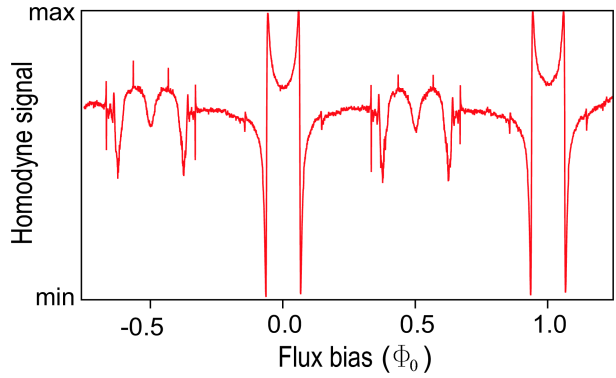}
\caption{Modulation of the reflected $8.18$~$\mathrm{GHz}$
microwave signal with externally applied flux $\Phi _{\mathrm{ext}}$. The
signal is clearly flux-periodic indicating that the junction ring is closed
and superconducting. The values of $\Phi _{\mathrm{ext}}$ at which the
signal undergoes full swings correspond to the anticrossings of the $0-1$
transition frequency of the device with the resonator bare frequency, later
inferred to be $8.1755$~$\mathrm{GHz}$. The measurement tone populates the
resonator with less than $0.01$ photon on average.}
\label{Fig.2}
\end{figure}

To characterize the fluxonium, we first measure the ground state resonator
pull as a function of $\Phi _{\mathrm{ext}}$. The results (Fig.\ 2) show the
expected $\Phi _{0}-$ periodicity as well as the avoided crossings of the
resonator frequency and the ground to excited state transitions. This
confirms that the entire $44$ junction loop is superconducting and that the
resonator-atom system is in the strong coupling regime of cavity QED~\cite%
{RaymondRMP}.

\begin{figure}[tbp]
\centering
\includegraphics[width = 1.0\columnwidth]{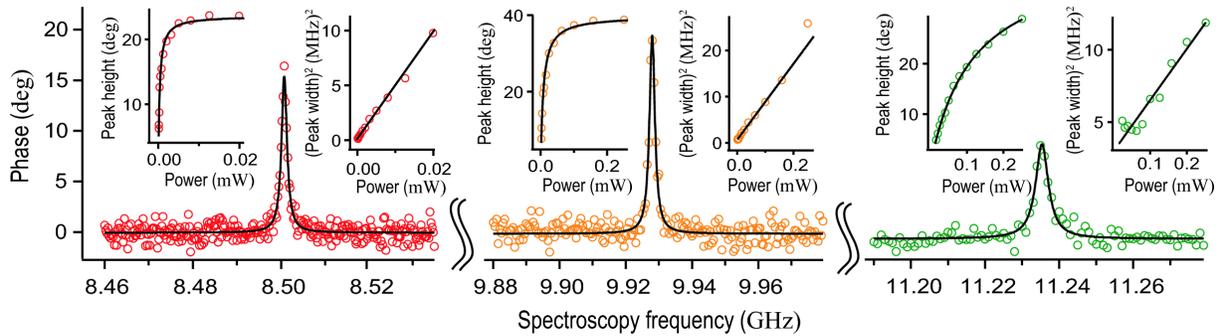}
\caption{Phase (colored circles) of reflected readout tone as a
function of spectroscopy tone frequency taken at $\Phi _{\mathrm{ext}%
}=0.05\Phi _{0}$. Data for the first three resonances (further identified as
transitions from the ground state to states $1$, $2$ and $3$) is shown from
left to right in red, orange and green respectively. Resonances are
well-fitted by Lorentzians (solid black lines) for a broad range of
spectroscopy powers. Insets on the two sides of each resonance show the
dependence of the resonant peak height (left) and width squared (right) on
the spectroscopy tone power. Data in all insets follow the predictions
(solid black lines) of Bloch equations describing relaxation dynamics for a
spin $1/2$ and indicate that all transitions involve one photon.}
\label{fig:Fig3}
\end{figure}

Next, we perform a two-tone spectroscopy measurement \cite%
{SchusterSpectroscopy} at a fixed flux $\Phi _{\mathrm{ext}}=0.05\Phi _{0}$,
during which, in addition to the fixed frequency readout tone, we probe the
transition frequencies of the atom through a second, variable frequency
spectroscopy tone. The resulting peaks (Fig.\ 3), correspond to the
later-determined $0-1$, $0-2$, and $0-3$ transitions from the atom ground
state. The peaks are well-fitted by Lorentzians and their power-dependent
widths and heights are well-explained by the Bloch equations of precessing
spin $1/2$~\cite{AbragamNMRbook} as shown in the insets of (Fig.\ 3).
Extrapolating fitted linewidths to zero spectroscopy power, we obtain lower
bound estimates of their decoherence time at $350$, $250$ and $80$~$\mathrm{%
ns}$ respectively.

\begin{figure}[tbp]
\centering
\includegraphics[width = 1.0\columnwidth]{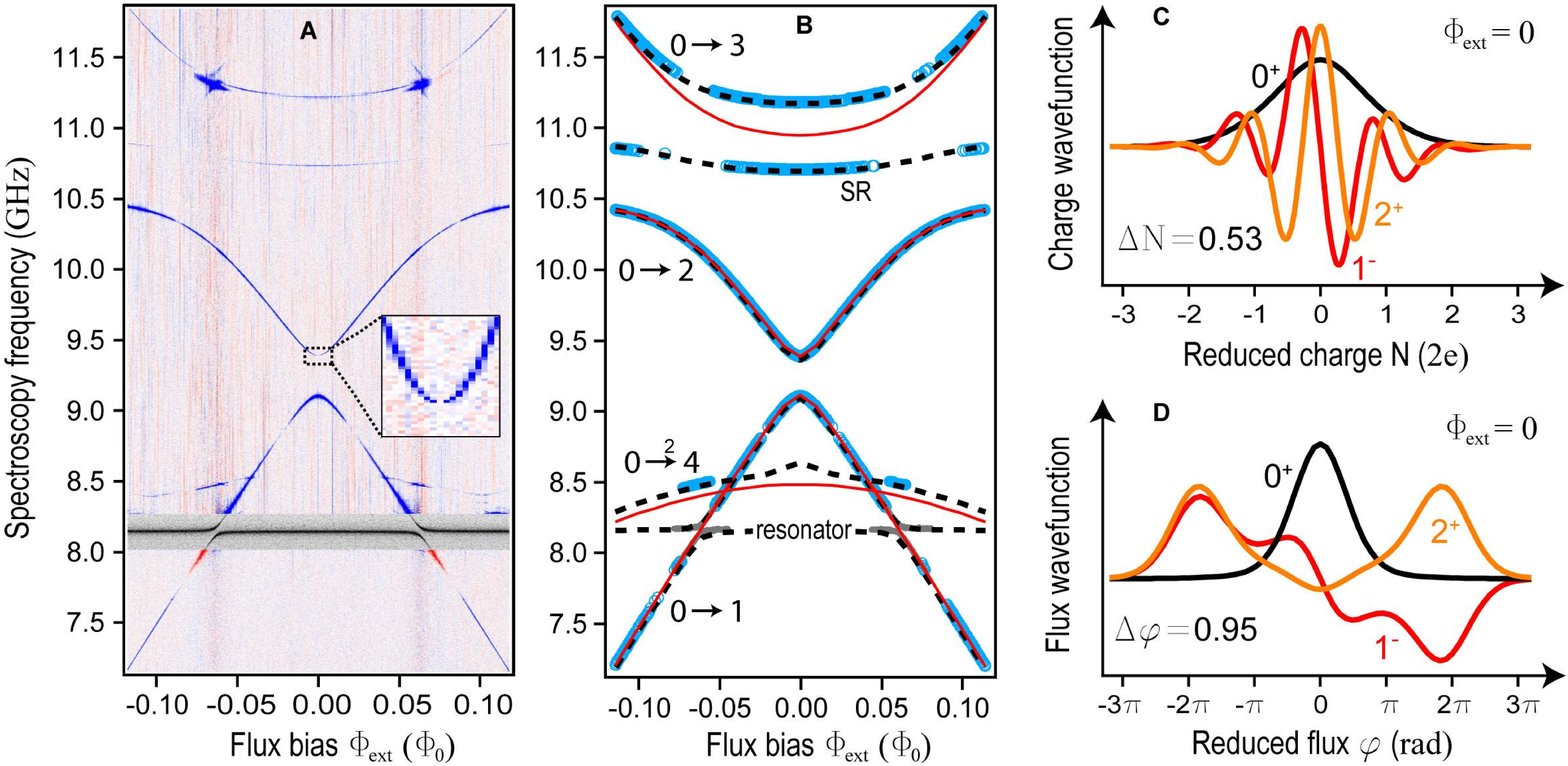}
\caption{(A) Phase of reflected readout tone as a function of
the spectroscopy tone frequency and external flux. The color scale encodes
the value of phase, zero corresponding to the mauve background, blue to
positive values (peaks) and red to negative values (dips). The grey region
shows the reflected phase of a single tone, swept close to the resonator
bare frequency exhibiting a $50~\mathrm{MHz}$ vacuum Rabi splitting of the
resonator with the fluxonium transition $0-1$. The inset in (A) zooms in the
central region of the $0-2$ transition and confirms that it is indeed
symmetry-forbidden at $\Phi _{\mathrm{ext}}=0$. (B) Measured peak
frequencies (blue circles) fitted by numerically computed spectrum of
Hamiltonian (2) (solid red lines) and its modification (see supplementary
online text) to explain the additional transition labeled \textquotedblleft
SR\textquotedblright\ (dashed black lines). (C) Amplitude of fluxonium wave
functions for levels $0~$(black), $1~$(red) and $2~$(orange) computed in
charge representation at zero flux bias, using circuit parameters extracted
from the fits. (D) Same as in (C) but in flux representation. The flux
representation wave functions demonstrate that the reduced flux is
delocalized compared to the size of the Josephson well while charge wave
functions confirm that the localization of charge on the junction is less
than a single Cooper pair charge. Note that in this circuit, the junction
charge is a continuous variable, in contrast to the Cooper pair box, and
flux swings of more than $2\pi $ are allowed.}
\label{fig:Fig4}
\end{figure}

Our main result is the spectroscopic data collected as a function of both
spectroscopy frequency and flux (Fig.\ 4A). Note that $\Phi _{\mathrm{ext}}$
variations span $20\%$ of $\Phi _{0}$ around $\Phi _{\mathrm{ext}}=0$
instead of the usual $1\%$ or less around $\Phi _{0}/2$ in flux qubit
experiments~\cite{ChargeFreeFluxQubit}. In (Fig.\ 4B) we compare the
measured peak center frequencies with the prediction for the $0-1$, $0-2$, $%
0-3$ and the two-photon $0-4$ transitions obtained from numerical
diagonalization of Hamiltonian (Eq. 2). Note that we are in effect fitting
more than three flux dependent functions, i.e. the flux dependent transition
frequencies,\emph{\ }with only three a priori unknown energies $E_{C}$, $%
E_{L}$ and $E_{J}$ so the problem is severely overconstrained. The fit of
the line (Fig.\ 4B) labeled SR (for array self-resonance) requires a minor
extension of the model taking into account parasitic capacitances across the
array \cite{MaterialsAndMethods}. Apart from introducing another resonator
mode coupled to the atom, this extension by no means invalidates the
inductive character of the array, at least as far as the $0-1$ and $0-2$
transition of the fluxonium are concerned. Even the perturbation of the $0-3$
and $0-4$ transition frequencies by this extra mode is less than $2\%$.

Based on the excellent agreement between theory and experiment, we infer the
wavefunctions of the first three energy levels, and plot their amplitudes
both in charge (Fig.\ 4C) and flux (Fig.\ 4D) representations for $\Phi _{%
\mathrm{ext}}=0$. In the ground state, we find that the ratio of charge to
flux fluctuations is $\Delta N/\Delta \varphi =0.56$, about $5$ times
smaller than the fine structure constant allows for a conventional
resonator. This confirms that the charge in our circuit is indeed localized
at the single Cooper pair level~($\Delta N=0.53$, $\Delta \varphi =0.95$).
The wavefunctions in flux representation (Fig.\ 4D) can be interpreted as
simple superpositions of states in which the reduced flux $\varphi $ is
localized in the wells of the Josephson cosine potential (fluxon states,
hence the name fluxonium). The parity of fluxonium states, which forbids the 
$0-2$ transition at zero external flux, manifests itself explicitly by a
remarkable ``hole" in the corresponding spectroscopic line (Fig.\ 4A,
inset). The allowed transition between the second and third level (data not
shown) is particularly spectacular since it corresponds to motion of the
total flux in the fluxonium loop by two whole flux quanta. This is to be
contrasted with the $10\%$ of flux quantum or less flux motion involved in
transitions of the flux and phase qubits~\cite%
{ChargeFreePhaseQubit,ChargeFreeFluxQubit}. Nevertheless, despite the large
flux fluctuations of the system and the corresponding charge pinning, the
circuit has complete immunity to offset charge variations: the data of
(Fig.\ 4A) has been taken piecemeal in $72$ hours and no jumps or drifts
have been observed during this period.

We have thus demonstrated that an array of Josephson junctions with
appropriately chosen parameters can perform two functions simultaneously:
short-circuit the offset charge variations of a small junction and protect
the strong non-linearity of its Josephson inductance from quantum
fluctuations. The data shows that the array possesses a microwave inductance 
$10^{4}$ times larger than the geometric inductance of a wire of the same $%
20~\mathrm{\mu m}$ length. The reactance of such inductor is about $%
3R_{Q}\approx 20$~$\mathrm{k\Omega }$ at $10~\mathrm{GHz}$ while its
resistance is less than $1~\Omega $. The spectrum of the fluxonium qubit
suggests it is as anharmonic as the flux qubit but as insensitive to flux
variations as the transmon qubit. Possible applications of this single
Cooper pair charging effect immune to charge noise include the observation
of fully developed macroscopic quantum coherent oscillations between fluxon
states~\cite{LeggetMQC}, the search for a \textquotedblleft $\mathrm{\Lambda 
}$\textquotedblright\ or \textquotedblleft $\mathrm{V}$\textquotedblright\
transition configurations for the shelving of quantum information~\cite%
{Shelving} in superconducting artificial atoms, topological protection of
superconducting qubits~\cite{KitaevTopProtQubit}, and finally the
long-sought quantum metrology of electrical current via Bloch oscillations~%
\cite{AverinBO, LikharevBO}.

\begin{scilastnote}

\item We acknowledge discussions with Markus Brink, Etienne Boaknin, Michael
Metcalfe, R. Vijay, David Schuster, Leo DiCarlo, Luigi Frunzio, Robert
Schoelkopf and Steven Girvin. This research was supported by the NSF under
grants DMR-0754613, DMR-032-5580, the NSA through ARO Grant No.
W911NF-05-01-0365, the Keck foundation, and Agence Nationale pour la
Recherche under grant ANR07-CEXC-003. M.H.D. acknowledges partial support
from College de France.
\end{scilastnote}

%%%%%%%%%%%%%%%%%%%%%%%%%%%%%%%%%%%%%%%%%%%%%%%%%%%%%%%%%%%%%%%%%%%%%%%%%%%%%%%
%%%%%%%%%%%%%%%%%%%%%%%%%%%%%%%%%%%%%%%%%%%%%%%%%%%%%%%%%%%%%%%%%%%%%%%%%%%%%%%
%%%%%%%%%%%%%%%%%%%%%%%%%%%%%%%%%%%%%%%%%%%%%%%%%%%%%%%%%%%%%%%%%%%%%%%%%%%%%%%
%%%%%%%%%%%%%%%%%%%%%%%%%%%%%%%%%%%%%%%%%%%%%%%%%%%%%%%%%%%%%%%%%%%%%%%%%%%%%%%

\newpage
\title{\Huge\textbf{Supporting Material}}

\maketitle

\section{Materials and Methods}

\textit{Sample fabrication.} The device is made on a high-resistivity Si substrate, $300~\mathrm{%
\mu m}$ thick. Both Josephson junctions and the readout resonator are
fabricated in a single step using e-beam lithography, double angle
Al e-beam evaporation and lift-off techniques. The Al evaporation and oxidation is
conducted in an e-gun evaporator at pressures less than $10^{-5}$ $\mathrm{Pa%
}$, AlOx grown in the environment of $680~\mathrm{Pa}$ of $15\%$ oxigen-in-argon
mixture during $10$ minutes. The areas of the small junction and array
junctions are designed to be $0.2\times 0.3~\mathrm{\mu m}^{2}$ and $%
0.25\times 2~\mathrm{\mu m}^{2}$, respectively. All $43$ array junctions are
equally spaced at less than $200$ $\mathrm{nm}$ so that total length of the array
is only $20~\mathrm{\mu m}$. The loop area of the array-small junction ring
is $3\times 20~\mathrm{\mu m}^{2}$.

\textit{Sample mount.} The Si chip is glued using GE varnish to the copper
base of a fully enclosing, custom-made microwave sample holder, shielding the
sample from both residual RF, infrared and optical photons. The holder
provides two well-matched transitions from the Anritsu K-connectors on the
outside of the holder to the two microstrip lines made on a PCB inside the
holder. The resonator's on-chip launching pads, schematically indicated as
two sections of a coaxial cable and marked \textquotedblleft $50~\mathrm{%
\Omega }$\textquotedblright\ in (Fig.\ 1A), are then wirebonded to the ends of
the two microstip lines.

\textit{Cryogenic setup.} The experiment is performed in a dilution
refrigerator with base temperature $10-20~\mathrm{mK}$. Both resonator and
the qubit are differentially excited via the $\Delta $-port of a $180^{\circ}$
 hybrid (Krytar, $2-18~\mathrm{GHz}$), whose two outputs are connected
to the two ports of the sample holder. Incoming and outgoing signals are
separated with a directional coupler (Krytar, $2-20~\mathrm{GHz}$). The
incoming signal line is attenuated using $10$ and $20~\mathrm{dB}$ microwave
attenuators (XMA) at all temperature stages of the refrigerator, to remove
non-equilibrium noise. The output line is amplified at the $4~\mathrm{K}$
stage with a low-noise HEMT amplifier (Caltech, $1-12~\mathrm{GHz}$, $30~%
\mathrm{dB}$ gain). Two cryogenic isolators (Pamtech, $4-12$~$\mathrm{GHz}$%
, $15~\mathrm{dB}$) are placed between the amplifier and the sample, at the $%
800~\mathrm{mK}$ stage and at the base stage, again to remove
non-equilibrium noise, especially that coming from the amplifier. Stainless
steel SMA cables are used to connect between the different temperature
stages. All components are thermally anchored to the proper refrigerator
stages. A  $\sim 1~\mathrm{cm}$ diameter custom made superconducting coil is
glued to the sample holder, a few $\mathrm{mm}$ away from the chip, to
provide perpendicular magnetic flux bias. The sample holder together with
the coil is placed into a Cryoperm cylinder to shield it from stray
quasistatic magnetic fields.

\textit{Room temperature measurement setup.} The readout resonator is
excited with Agilent E8257D signal generator, the spectroscopy signal is
generated using Agilent E8267D vector signal generator and Tektronix 520
AWG. Both signals are combined at room temperature and sent into the input
line of the refrigerator. The reflected $\sim 8~\mathrm{GHz~}$readout signal
from the refrigerator output line is amplified at room temperature with two
Miteq amplifiers ($1-12~\mathrm{GHz}$, $30~\mathrm{dB}$ gain), mixed down
with a local oscillator (a third Agilent E8257D) to an IF signal of $0-50~%
\mathrm{MHz}$, filtered and amplified with the IF amplifier (SRS SR445A),
and finally digitized using $1~\mathrm{GS}/\mathrm{s}$ Agilent Acqiris
digitizer. A software procedure then extracts the phase and the amplitude of
the digitized wave. The experiment is typically repeated $10^{4}$ times to
average the Gaussian noise to an acceptable level. Because the duration of
each experiment is about $10$ microseconds, every averaged data point is
taken in a fraction of a second. All microwave test equipment is phase
locked using a Rb precision $10~\mathrm{MHz}$ reference (SRS FS725). The
magnetic coil is biased in series with a resistor with Yokogawa 7751 voltage
source.

\textit{Comments on the data.} The data in (Fig.\ 2) shows the digitized
homodyne (zero IF) signal as a function of magnetic field, with the
spectroscopy generator turned off. The data in (Fig.\ 3) shows the phase of
the digitized heterodyne ($50$ $\mathrm{MHz}$ IF) signal, as a function of
frequency of the spectroscopy generator. The data in (Fig.\ 4A) is taken in
the pulsed regime, when the spectroscopy generator outputs a $6~\mathrm{\mu s%
}$ saturating pulse followed immediately by the $2~\mathrm{\mu s}$ readout
pulse. This way we ensure that the sample is exposed to only one tone at a
time, avoiding various spurious effects. The image presented in (Fig.\ 4A) contains 
$367\times4597$ data points.

\section{Supplementary Text}
\begin{figure}
	%\centering
	  %\psfrag{t}[l][][1.5]{$\Phi_\mathrm{ext}$}	
	  %\psfrag{f}[l][][1.5]{$\varphi$}	
		\includegraphics[width=0.8\textwidth]{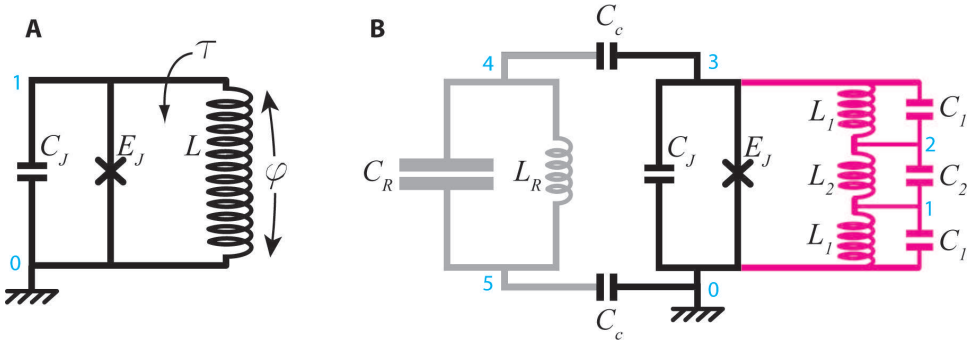}
	\caption{Models for the fluxionium device. (A) Inductively shunted junction model. (B) Extended fluxionium model, including the capacitive coupling to the mode of a transmission-line resonator, and parasitic capacitances across the array. Numbers in cyan enumerate the nodes of the circuits.\label{fig:circuit}}
\end{figure}

In our analysis of the fluxonium device, we use two simple models whose corresponding circuits are depicted in Figure \ref{fig:circuit}: (A) the inductively shunted junction model, (B) the extended fluxonium model, describing the fluxonium coupled to a transmission-line resonator.

\textit{Inductively shunted junction model}.
The simplest model of the fluxonium device is the inductively shunted junction model, see Fig.\ \ref{fig:circuit}(A). It neglects parasitic capacitances across the fluxonium's Josephson junction array, and assumes that all internal degrees of freedom of the array are frozen out. In this limit, the fluxonium consists of a single Josephson junction with capacitance $C_J$ and Josephson energy $E_J$, shunted by a large inductance $L$. Quantization of this circuit \cite{DevoretQFinEC} is straightforward and leads to the Hamiltonian
\begin{equation}
\hat{H}_0=4E_C \hat{N}^2 +\frac{1}{2}E_L\hat{\varphi}^2 -E_J\cos\left(\hat{\varphi}-2\pi\frac{\Phi_\mathrm{ext}}{\Phi_0}\right),
\end{equation}
where the charge on the junction capacitance $\hat{N}$ (in units of $2e$) and the reduced flux $\hat{\varphi}$ are canonically conjugate variables, $[\hat{\varphi},\hat{N}]=i$. Structurally, this Hamiltonian is identical to the Hamiltonian describing one-junction flux qubits, and flux-biased phase qubits. However, the regime of large inductances relevant for the fluxionium differs from typical parameters in flux and phase qubits, and has been discussed in Ref.\ \cite{JensCPBL}.

\textit{Extended fluxonium model}.
For a more complete modelling of the spectra obtained in the experiment, we take into account the coupling of the fluxonium device to a transmission-line resonator. In addition to this resonant mode, the experimental data shows another resonance coupling to the fluxonium device. Such additional resonances are expected when accounting for the parasitic capacitances of the Josephson junction array. The simplest effective model accurately describing the experimental data includes very few of these capacitances, and approximates the array by a combination of inductances and capacitances as shown in Fig.\ \ref{fig:circuit}(B).

The Lagrangian describing this circuit can be written in the form
\begin{eqnarray}
\mathcal{L}=&\frac{C_J'}{2}\dot{\phi}_3 - \frac{1}{2L}(\phi_3-\Phi_\mathrm{ext})^2+E_J\cos\left( \frac{2\pi\phi_3}{\Phi_0} \right)+C'\dot\theta_2-\frac{1}{L'}\theta_2^2
+\frac{C_R'}{2}\dot\varphi_4^2-\frac{1}{2L_R}\varphi_4^2\\\nonumber
&+\frac{C_c}{2}\dot\phi_3\dot\varphi_4 +\tilde{C}\dot\theta_2\dot\phi_3,
\end{eqnarray}
where $C_J'=C_c/2+C_J+2C_1\lambda_1^2+C_2\lambda_2^2$, $\lambda_i=L_i/L$, $C'=9(C_1+2C_2)\lambda_1^2\lambda_2^2$, $L'=L/(9\lambda_1\lambda_2)$, $C_R'=C_R+C_c/2$, $\tilde{C}=6\lambda_1\lambda_2(C_1\lambda_1-C_2\lambda_2)$, and we have disposed of another resonant mode which does not couple to the fluxonium device. In terms of the original generalized flux $\phi_i$ at each node $i$, the relevant variables are $\phi_3$ (associated with the fluxonium subsystem), the resonator mode $\varphi_4=\phi_4-\phi_5$, and the additional resonant mode $\theta_2=-\frac{1}{6\lambda_1\lambda_2}(\phi_2-\phi_1-\lambda_2\phi_3)$. Employing canonical quantization of this circuit, we find the effective Hamiltonian
\begin{equation}
\hat{H}=\hat{H}_0 + \sum_{j=1,2}\hbar\omega_j \hat{a}_j^\dag \hat{a}_j +\hbar\sum_{j=1,2}g_j\hat{N}(\hat{a}_j^\dag + \hat{a}_j),
\end{equation}
describing the inductively shunted junction, $\hat{H}_0$, coupled to two resonant modes $j=1,2$ with coupling strengths $g_1$ and $g_2$, respectively.

\textit{Theory fits to experimental data}.
Design and fabrication of the fluxonium system only allow for imprecise estimates of the system parameters. Thus, the comparison between experimental data and theoretical prediction requires the fitting of theory curves to determine the system parameters with more accuracy. The parameters at our disposal are: $E_J$, $E_C$, and $E_L$ (for both the inductively shunted junction and the extended fluxonium models). In addition, the extended fluxonium model takes the resonant mode frequencies and coupling strengths $\omega_{1,2}$ and $g_{1,2}$ as input.
Fits are obtained by extracting the center frequencies from the experimental data and employing a least-squares fit algorithm. 

\textit{Fit to inductively shunted junction model}.
A simultaneous fit to the full flux-dependence of the 0--1 and 0--2 transitions around the zero-flux point fully determines the fluxonium parameters $E_C$, $E_J$, and $E_L$ (see Table \ref{tab} for the obtained parameter values). A comparison between the resulting theory prediction of higher transitions can then be used as a consistency check. While the agreement for the 0--1 and 0--2 transitions is good, we find systematic deviations for higher levels. The reason for these deviations lies in the effect of the additional resonance on the 0--2 transition: the additional resonance leads to significant frequency shifts of the 0--2 transitions. Ignoring this effect leads to a systematic error in the estimation of the fluxonium parameters.

\textit{Fit to extended fluxonium model}.
For best agreement, both resonator and additional resonant mode are taken into account.  Using the full experimental data we obtain a fit for the extended fluxonium model, which shows excellent agreement with the data. The resulting parameter values are given in Table \ref{tab}.
\begin{table}
	\centering
		\begin{tabular}{lcc}\hline\hline
		& inductively shunted & \hspace*{0.5cm}extended fluxonium model\\
		& junction model      & \\\hline
$E_C/h $& 2.39 & 2.47\\
$E_J/h $& 8.93 & 8.97\\
$E_L/h $& 0.52 & 0.52\\
$\omega_1/2\pi $& n.a. & 8.18\\
$\omega_2/2\pi $& n.a.& 10.78\\
$g_1/2\pi  $& n.a. & 0.135\\
$g_2/2\pi  $& n.a. & 0.324\\\hline\hline
		\end{tabular}
		\caption{Fluxonium system parameters obtained from least-squares fits to the inductively shunted junction and the extended fluxonium model. All values are given in GHz. The coupling constants are expressed in terms of the coupling strength for the fluxonium 0--1 transition.\label{tab}}
\end{table}

%%%%%%%%%%%%%%%%%%%%%%%%%%%


\begin{thebibliography}{99}
\bibitem{FineStructureNote} Note that the ratio $(\mu _{0}/\varepsilon
_{0})^{1/2}/R_{Q}=16\pi \alpha $, where $\alpha \approx 1/137$ is the fine
structure constant.
\bibitem{MaterialsAndMethods} Materials and methods are available as
supporting material on Science online.
\bibitem{theEcfootnote} Due to coupling of the fluxonium circuit to the
readout resonator (Fig.1D) $C_{J}$ is renormalized to the value $C_{J}+C_{c}$%
, with corrections of order $C_c/C_R\simeq1\%$.
\bibitem{footnote3} Note that while the circuit in (Fig.\ 1D) might appear
identical to the RF-SQUID~circuit~\cite{RF-SQUID}, there is a drastic
difference between the two when charging, Josephson and inductive energies
are compared. In addition, RF-SQUID is addressed by coupling to its flux
degree of freedom via mutual inductance, while the fluxonium is addressed by
coupling to its charge degree of freedom via capacitance $C_{c}$.
\end{thebibliography}
\end{document}